\documentclass[journal,article,preprints,submit,pdftex,moreauthors]{Definitions/mdpi} 

\firstpage{1} 
\pubvolume{1}
\issuenum{1}
\articlenumber{0}
\pubyear{2024}
\copyrightyear{2024}
\datereceived{ } 
\daterevised{ } % Comment out if no revised date
\dateaccepted{ } 
\datepublished{ } 
\hreflink{https://doi.org/} % If needed use \linebreak
\Title{Measuring and Analyzing Subjective Uncertainty in Scientific Communications}

\Author{Grace Shao $^1$ and Jamshid Sourati $^{2}$*\orcidB{}}

\AuthorNames{Grace Shao and Jamshid Sourati}

\address{%
$^{1}$ \quad University of Chicago, Department of Sociology (Computational Social Sciences)\\
$^{2}$ \quad DePaul University, School of Computing}

\corres{Correspondence: jsourati@depaul.edu}

\abstract{Uncertainty of scientific findings are typically reported through statistical metrics such as $p$-values, confidence intervals, etc. The magnitude of this objective uncertainty is reflected in the language used by the authors to report their findings primarily through expressions carrying uncertainty-inducing terms or phrases. This language uncertainty is a subjective concept and is highly dependent on the writing style of the authors. There is evidence that such subjective uncertainty influences the impact of science on public audience. In this work, we turned our focus to scientists themselves, and measured/analyzed the subjective uncertainty and its impact within scientific communities across different disciplines. We showed that the level of this type of uncertainty varies significantly across different fields, years of publication and geographical locations. We also studied the correlation between subjective uncertainty and several bibliographical metrics, such as number/gender of authors, centrality of the field’s community, citation count, etc. The underlying patterns identified in this work are useful in identification and documentation of linguistic norms in scientific communication in different communities/societies.}


\begin{document}

%%%%%%%%%%%%%%%%%%%%%%%%%%%%%%%%%%%%%%%%%%
%\setcounter{section}{-1} %% Remove this when starting to work on the template.
%\endnote{This is an endnote.} % To use endnotes, please un-comment \printendnotes below (before References). Only journal Laws uses \footnote.

% The order of the section titles is different for some journals. Please refer to the "Instructions for Authors” on the journal homepage.

\section{Introduction}

Traditionally, scientific theories were taken as absolute claims with no subjective uncertainty around them. Today, we recognize that our scientific models are merely computational approximations of reality and, therefore, are taken within a shade of uncertainty as opposed to absolute assertions~\cite{kampourakis2019uncertainty}. Uncertainty of findings could be either obtained through the statistical analysis reported in the manuscripts through $p$-values, confidence intervals, etc.~\cite{boschen2021evaluation, kalmbach2023rule}, and/or directly based on internal structure of the findings themselves~\cite{yu2022uncertainty, svensson2024enhancing}\footnote{Such quantifiable objective uncertainty is sometimes called ``risk'' against the non-quantifiable lack of confidence that is simply called ``uncertainty''~\cite{knight1921risk}.}. The formulations used to measure this type of quantifiable uncertainty can be derived independently of the scientists who conducted the work, which is why it is regarded as \emph{objective} uncertainty. The degree of objective uncertainty is typically reflected in the language used by the authors to convey their findings primarily through expressions that incorporate uncertainty-inducing terms or phrases. While objective uncertainty can be reported through statistical metrics independent of the authors' writing style, the verbal uncertainty is a \emph{subjective} concept depending on how individual writers communicate their findings~\cite{dhami2005interpersonal}. Being able to properly interpret subjective uncertainty within the language can be helpful in the absence of reported statistical uncertainty analysis. Prior research provided evidence that the level of subjective uncertainty influences the impact of science on public audience. As an example, there have been studies showing destructive effect of high uncertainty on public trustworthiness~\cite{osman2018problems}, slowing down the dissemination of scientific results through social media~\cite{stavrova2024expressions}. On the other hand, it is suggested that very high certainty can damage the authors' reputation~\cite{dries2024evidence}. In this paper, we switched our focus on the effect of the way scientists incorporate uncertainty in their language on their peer researchers that are potentially from other fields.

Effectively communicating subjective uncertainty with respect to scientific findings is essential not only for the public audience but also for peer scientists and academicians. Understanding how to interpret findings of other researchers, especially across different disciplines, is vital for efficient knowledge exchange and interdisciplinary collaboration. This understanding will be improved by gaining insight into uncertainty-relevant language norms in different fields. It is suggested that academic institutions train researchers to effectively communicate their uncertainty to ensure adherence to the highest academic standards when imparting their findings~\cite{morgan2024communicating}. A question that arises is whether we have a unique set of standards for language use throughout science. Is a statement regarded as an uncertain assertion in "Physics"  also perceived as an uncertain claim in "Sociology" (disregarding the contents), and vice versa? 
%Moreover, quantified subjective uncertainties could serve as a textual indicator suggesting the rise of a paradigm shift.

% importance of measuring language uncertainty
There is evidence from previous studies implying that different aspects of language norms differ among different scientific disciplines~\cite{kjersti2006academic}. There is no standard English to which the submitted scientific works are compared to in the review process, apart from a typical English grammar check. As a result, the use of English in scientific publications can significantly vary across different fields, different countries and even different times~\cite{netzel2003way}.  Such variations can lead to misinterpretations for both human and automated readers. One approach to mitigating this issue is to identify and document linguistic norms in scientific communication of different communities/societies. There are studies analyzing different linguistic aspects of academic papers including language informality features~\cite{hyland2017academic}, positive words~\cite{vinkers2015use, edlinger2023presence} and nominalizations~\cite{biber2021nominalizing}.
%For example, Hyland and Jiang~\cite{hyland2017academic} showed that while the language of academic papers in Sociology and Applied Linguistic has become more formal during a recent period of 20 years, Electrical Engineering and Biology observed a significant increase in the use of informality features.
%This type of uncertainty have been shown to have significant influence on audiences' perception~\cite{gustafson2020review}. 
Additionally, given that not all papers explicitly report objective uncertainty measurements with their findings, measuring subjective uncertainty, though noisy and subjective, can help us better calibrate and finetune our text-based objective uncertainty estimators~\cite{ningrum2025annotating}.

Here, we focused on the subjective uncertainty aspect of science communication along three dimensions: disciplines, time of publications and geographical location of institute affiliations. The topic-wise and geographical differences can complicate interdisciplinary communications posing a barrier to collaborative efforts. Whereas the temporal variations can pose difficulty when fully comprehending the studies published by previous generations of researchers. In this paper, we studied the uncertainty language norm in relationship to a range of bibliographical metrics. Our research in this work involved measuring and analyzing verbal uncertainty present in scientific articles irrespective of the underlying source of lack of confidence that may have caused the uncertainty.

Technically speaking, there have been at least three types of methodologies used for quantifying subjective uncertainty: machine learning models that are fine-tuned specifically towards measuring uncertainty of any given sentence~\cite{prieto2020data, pei2021measuring}; machine learning models that are trained for another purpose but whose embeddings could be used for measuring subjective confidence in a sentence; and finally approaches that have no learning parameters, e.g., algorithms based on hedging and modality detection~\cite{hyland1996writing, vincze2013weasels, pyatkin2021possible}. In our work, we first made a comparison between the available uncertainty measurement tools and used the one with the most reliable result based on our own annotated dataset. 

%The introduction should briefly place the study in a broad context and highlight why it is important. It should define the purpose of the work and its significance. The current state of the research field should be reviewed carefully and key publications cited. Please highlight controversial and diverging hypotheses when necessary. Finally, briefly mention the main aim of the work and highlight the principal conclusions. As far as possible, please keep the introduction comprehensible to scientists outside your particular field of research. Citing a journal paper \cite{ref-journal}. Now citing a book reference \cite{ref-book1,ref-book2} or other reference types \cite{ref-unpublish,ref-communication,ref-proceeding}. Please use the command \citep{ref-thesis,ref-url} for the following MDPI journals, which use author--date citation: Administrative Sciences, Arts, Econometrics, Economies, Genealogy, Humanities, IJFS, Journal of Intelligence, Journalism and Media, JRFM, Languages, Laws, Religions, Risks, Social Sciences, Literature.

%%%%%%%%%%%%%%%%%%%%%%%%%%%%%%%%%%%%%%%%%%
\section{Materials and Methods}

\subsection{Dataset}

The main dataset used for this study consisted of a subset of Microsoft Academic Graph (MAG)~\cite{sinha2015overview}, a database of millions of scientific publications up to the end of year 2021. With each publication is saved its title, abstract, author names and more metadata (e.g., publication year, author affiliations, etc.). We chose to work with publications in 10 disciplines listed in Table~\ref{tab:fields}, ignoring entries in the database that lack an abstract, or their abstracts are not in English. The main fields of any given paper are determined based on the level-0 tags in MAG. We classified sentences of each abstract into 7 classes according to the method proposed by~\cite{prabhakaran2016predicting} and selected sentences classified as the ``conclusion''. On average, about 17.4\% papers had English abstracts with at least one conclusive sentence, with Biology showing the highest percentage (31\%) and Psychology showing the lowest fraction (8.8\%).

\begin{table}[H] 
\caption{Statistics of the ten scientific fields selected from MAG database. It shows the total number of papers, in column ``\#papers'', number of papers with English abstracts, in column ``\#Papers w/ Abstracts'', and number of abstracts with at least one conclusive sentences, in column ``\#Abstracts w/ Conclusion'' (M$=\times 10^6$, K$=\times10^3$). Rows are sorted in an descending order in terms of the last column. \label{tab:fields}}
\begin{tabularx}{\textwidth}{LCCC}
\toprule
\textbf{Field}	& \textbf{\#Papers}	& \textbf{\#Papers w/ Abstracts} & \textbf{\#Abstracts w/ Conclusion}\\
\midrule
Materials Science	& 31.7 M & 20.2M & 5.6M  \\
Biology	& 14.2M	& 8M & 4.4M  \\
Computer Science	& 28M	& 18M & 4.34M  \\
Chemistry	& 18.9M M & 10.5M & 3.9M  \\
Physics		& 9.4M	& 5.4M & 1.7M \\
Mathematics	& 7.2M	& 4.3M & 1.16M  \\
Electrical Engineering	& 5.9 M	& 4.23M & 1M  \\
Psychology	& 8.5M	& 3.7M & 748K  \\
Economics	& 3.2M	& 1.7M & 638K  \\
Sociology	& 4.6M	& 1.8M & 423K  \\
\bottomrule
\end{tabularx}
%\noindent{\footnotesize{\textsuperscript{1} Tables may have a footer.}}
\end{table}

\begin{table}[H] 
\caption{Distribution and examples of annotated sentences. \label{tab:annots}}
\begin{tabularx}{\textwidth}{LCL}
\toprule
\textbf{Label}	& \textbf{Size}	& \textbf{Eaxmple} \\
\midrule
Uncertain  &  237 & The action appears to be specific and mediated by the hTSHR.  \\
Neutral  & 662 &  This increase is first observed 4 d after injection of anti-IgD. \\
Certain  &  50  &  Our model accurately predicts the kinetics and extent of receptor downregulation induced by either EGF or TGF-alpha.\\
\bottomrule
\end{tabularx}
%\noindent{\footnotesize{\textsuperscript{1} Tables may have a footer.}}
\end{table}

We used a separate annotated dataset for evaluating the performance of different uncertainty measurement approaches considered in this study (explained in the next sub-section). To create this data, we manually labeled 949 sentences extracted from randomly selected scientific papers in life sciences topics in PubMed database. Following the analysis of~\cite{prieto2020data} on the level of certainty levels, we annotated our sentences to one of the three levels of certainty such that ``1'' indicates the least certainty (highest uncertainty) and ``3'' represents highest certainty (least uncertainty) (Table~\ref{tab:annots}). 

\subsection{Certainty Measurement}
A wide range of algorithms have been proposed to measure language uncertainty in a sentence or document. In this work, we considered five different strategies listed below. For more details, see Appendix~\ref{app:models}.
\begin{itemize}
    \item a SciBERT model that has been fine-tuned over a set of certainty-annotated scientific claims~\cite{pei2021measuring};
    \item a 5-layer Long Short Term Memory (LSTM) model trained on uncertainty annotations~\cite{prieto2020data};
    \item zero-shot classification of uncertain/certain sentences using a pre-trained large-BART transformer~\cite{lewis2020bart};
    \item using a generative AI model (GPT 4) for inquiring about the language uncertainty of the given texts~\cite{OpenAI2023ChatGPT};
    \item an unsupervised hedge-based model, where we count mentions of modifiers and clauses from a list of 540 hedges extracted based on biomedical articles and general texts~\cite{vincze2008bioscope, vincze2013weasels}.
\end{itemize}

Each of these methods returns a single, non-negative scalar score for a given sentence. The resulting outputs are linearly transferred to fall in the same range as our annotations ($\in [1,3]$) such that 1 indicates least certainty (i.e., highest uncertainty) and 3 represent highest certainty (i.e., least uncertainty). Therefore, the final scores primarily measured \emph{certainty} of a sentence. We used the annotated dataset mentioned above as the ground truth for evaluating the performance of certainty measurement approaches. Each model is applied on the same set of sentences and the degree of alignment between the resulting certainty scores and our annotations is quantified through correlation coefficient. The model with the highest alignment metric is selected and used as the winning method for all the remaining data analysis steps of this study.

Multiple scores would be obtained for papers with multiple conclusive sentences. We tried several summarization methods including averaging and miximization to report a single scalar per publication: 
\begin{equation}
    c(p) \enskip=\enskip \mbox{SUMM} \Big(c(s): s\in\mbox{CONC}_p \Big),
\end{equation}
where $p$ is a paper with sentences denoted by $s$ such that $\mbox{CONC}_p$ represents the conclusive sentences, $c$ is the certainty measurement function, and $\mbox{SUMM}$ is the summary function of our choice. Whereas the certainty of individual sentences $c(s)\in\mathbb{R}^+$ is taken as the output of one of the models above, the certainty of the whole paper $c(p)$ depends on our choice of the summary function $\mbox{SUMM}$. We tested two choices for $\mbox{SUMM}$: averaging ($\mathbb{E}$) or minimization ($\min$). Averaging the certainty scores tended to flat them out as there were usually very few uncertain sentences among many more neutral ones. We argue that the localized approach of taking the minimum certainty score is more suitable for our application, since having only a single (or very few) uncertain sentence(s) is enough to conclude that authors show lack of confidence about at least part of their findings, even though being surrounded by neutral or certain sentences.
% try to clarify this last sentence

\subsection{Certainty Analysis}
\begin{figure}[t]
\includegraphics[width=\columnwidth]{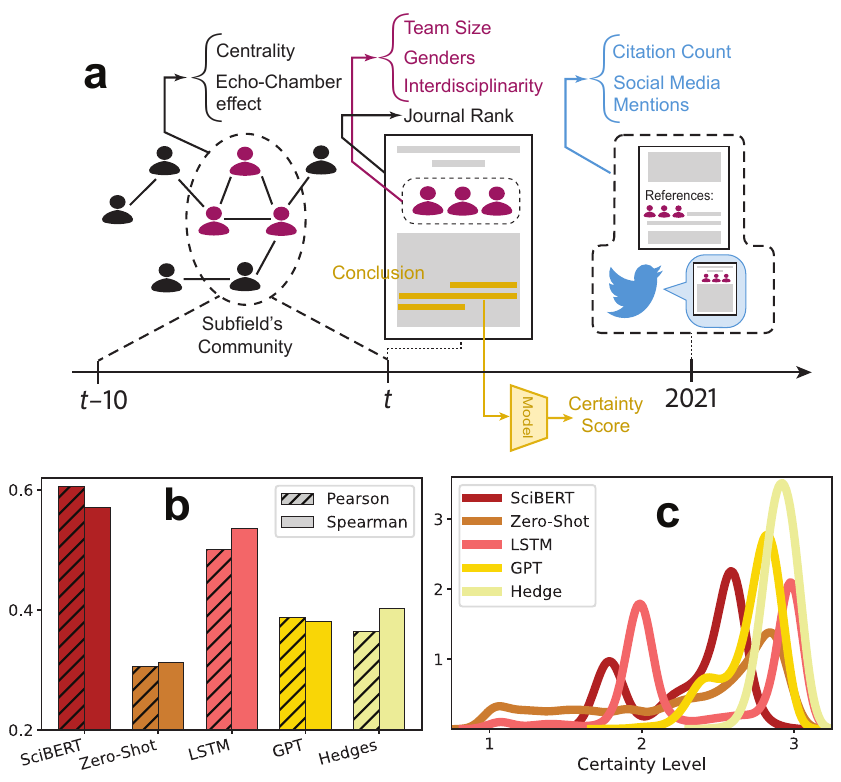}
\caption{(a) Three types of metrics used in our correlation analysis: pre-publication metrics describing the coauthorship graph of the community of a paper's subfield based on the most recent 10-year time window preceding the publication time; publication metrics characterizing the author teams as well as the publishing venue; post-publication metrics measuring impact of the paper in internal and external communities. (\textbf{b, c}) Performance of uncertainty measurement approaches we tried: (\textbf{a}) alignment of the methods' output with our annotations based on correlation coefficients. (\textbf{b}) Density estimates of the uncertainty scores of the competing methods on the evaluation dataset.
\label{fig:methods_perf}}
\end{figure}

Once we computed the certainty scores for all the involved publications, we ran our data analysis which included three main steps:
\begin{itemize}
    \item \textit{Temporal analysis:} First, we studied of the temporal pattern of certainty scores across different fields. In this study, we conducted a year-wise analysis, focusing solely on the publication year while disregarding the month and day of publication dates. Temporal patterns could be used as a sign for locating shifts in the norms of language use within scientific communities. %Secondly, to investigate whether our certainty measures could be associated with paradigm shifts (i.e., Kunian revolutions), we repeated the same temporal analysis on certain subfields that have arguably experienced a (psuedo–)paradigm shift relatively recently. 
    \item \textit{Correlation analysis:} Secondly, we studied any potential association between certainty measurements and several bibliographical metrics. We also conducted a temporal analysis in this section by examining patterns of correlations with respect to variety of metrics over time. As illustrated in Figure~\ref{fig:methods_perf}(a), one can categorize these metrics into publication criteria, as well as \emph{pre-} and \emph{post-}publications metrics (Table~\ref{tab:metrics}).
    \item \textit{Geographical Analysis:} Lastly, we investigated any geographical patterns in the measured certainty of the abstracts. we relied on the affiliation of authors for assigning geographical location to papers. We assigned the country of papers with multiple authors to the affiliation country of the first and last authors, as they typically contribute the most to the writing. Papers where the first and last authors are from different countries were excluded, ensuring that each paper is assigned to a single country.
\end{itemize}

\renewcommand{\arraystretch}{1.25}

\begin{table}[H] 
    \caption{Summary of metrics in correlation analysis. \label{tab:metrics}}
    \begin{tabularx}{\textwidth}{Lp{2in}p{1.6in}}
    \toprule
    \textbf{Metric}	& \textbf{Description}	& \textbf{Operationalization} \\
    \midrule
    \multicolumn{3}{c}{\small \textit{Pre-publication}}\\[-.1cm]
    \midrule
    Centrality	& degree of subfield domination by small number of authors & Gini coefficient of the coauthorship network \\
    Echo-Chamber Effect	& lack of exposure to other subfields & fraction of coauthorship connections between home and other subfields \\
    \midrule
    \multicolumn{3}{c}{\small \textit{Publication}}\\[-.1cm]
    \midrule
    Number of authors & number of researchers co-authoring a paper & \\
    Gender of authors	& probability that the first/last/all author(s) is/are male & logistic regression model\\
    Interdisciplinarity	& discipline diversity within the team of authors & dispersion of the team's distribution of disciplines  \\
    Journal rank & prestige of the publishing journal & journal rank record in MAG \\
    \midrule
    \multicolumn{3}{c}{\small \textit{Post-publication}}\\[-.1cm]
    \midrule
    Citation counts & number of papers cited the publications	& cite count metadata included in MAG \\
    Social media posts	& attention on social media	& $\#$ tweet mentions \\
    \bottomrule
    \end{tabularx}
    %\noindent{\footnotesize{\textsuperscript{1} Tables may have a footer.}}
\end{table}

In our correlation analysis, publication criteria include the number of researchers coauthoring a paper (hereafter \emph{authors}), gender of authors, interdisciplinarity of the authoring team, and prestige of the publishing venue. Our pre-publication metrics include factors characterizing the sub-field community to which the paper's main topic belongs, i.e., centrality and echo-chamber effect of the community's coauthorship network. Moreover, the post-publication criteria quantify impact of the paper's publication inside and outside the scientific communities, via the number of citation counts and the number of tweets that mentioned the paper, respectively. To measure the pre-publication criteria for any given paper, we built the community network of its sub-field based using literature from the 10 years preceding the paper's publication. On the other hand, when establishing the association between measured certainty and post-publication metrics, we only considered and compared the impacts of those papers published in the same year to control for temporal confounding factors. The internal and external impacts are measured via count of citations and mentions in social media, respectively. The criteria are explained below in more details (also see Appendix~\ref{app:bib_metrics}).

\subsubsection{Gender of Authors}
\label{subsubsec:gender}
Most of the bibliographic databases, such as MAG, do not contain gender information of the authors. To work around this missing information, we made gender inferences based on authors' first names through a pre-trained logistic regression model. This model operates on character n-grams as feature vectors, which enables the model to make inference for any names regardless of their existence in the training dataset. Specifically, we calculated the tf-idf of the character grams to be used as feature vectors. We normalized the resulting feature vectors through their log-ratios and trained a class-scaled logistic regression~\cite{hu2021s} over the Social Security Administration (SSA) database~\cite{social2018national}, which includes gender and popularity information of around 100,000 newborn names between years 1880 and 2018. Using an 80/20 train-test-split, the model gave a F$_1$ score of 0.7606 and an ROC of 0.8081. To label authors, the model was trained using the full SSA dataset and its probabilistic prediction, i.e., probability of being male for any author's first name, as the \emph{gender score} of that author. Authors for whom the first names are either lacking or abbreviated (e.g., ``C.'') are ignored.

We are interested in associating uncertainties to gender scores for individual papers. Hence, we need a mechanism for extending gender scores of individual authors to a whole paper when multiple authors exist. In the absence of metadata that indicates those with higher writing contributions, we have several alternatives to assign a single probabilistic score to the authors team. In this work, we tried the following three options when running our data analysis:
\begin{enumerate}[label=(\roman*)]
    \item extending gender score computed for the \emph{first} author to the whole paper--the assumption here is that the first author has the highest contribution in writing the paper.
    \item extending gender score computed for the \emph{last} author to the whole paper--the assumption here is that the last author has the highest contribution in writing the paper.
    \item assigning the average male probability computed for all authors to the whole paper--the assumption here is that all authors have almost the same writing contributions.
\end{enumerate}
These approaches have a degree of error due to the simplifying assumption on the level of the authors' writing contribution. This error increases as the number of authors in the publication increases. In order to keep this error upper-bounded, we limited our gender analysis to papers with fewer than 10 authors. 

\subsubsection{Interdisciplinarity of Teams}
Identifying interdisciplinarity publications could be done based on either content of the research, or academic records/affiliations of individuals who created that content. For example, \cite{xiang2024evaluating} distinguished two dimensions of interdisciplinarity: topic-base dimension which captures diversity of subject matters of the work, and knowledge-base category that identifies the diversity of ideas or knowledge retrieved from different disciplines and utilized to develop its arguments around the findings. Our current study pertains the subjective use of language made by authors in a publication. We introduce a third, author-based dimension for measuring interdisciplinarity of a publications, considering it as the outcome of collaboration among authors with diverse past experiences and backgrounds. In this method, we used subject matters of the previous publications to create an academic profile for each of the authors. Zero-level tags assigned to papers by MAG are used to specify the main topic(s) of any given paper. \emph{Research background} of a scientist is defined as the frequency of research topics extracted from his/her past publications. We normalized these frequencies and mathematically represented the author's research background by means of a probability mass function (PMF) across all 19 main topics defined in MAG. The interdisciplinarity of a team of authors could then be measured via dispersion of their research background PMFs. The dispersion metric is obtained by computing the exponentiated average of cosine distances between individual PMFs and their centroid: $\exp\left(\frac{-1}{n_a}\sum_{i}\cos(v_i,\bar{v})\right)$, where $v_i$ is the research PMF of the $i$-th author, and $\bar{v}$ is the centroid research background for a team with $n_a$ authors. Note that when $n_a=1$, the interdisciplinariness is zero by definition. On the other hand, the likelihood of having a higher dispersion increases when $n_a$ becomes larger.

\subsubsection{Centrality}
\label{subsubsec:centrality}
Centralized communities are groups of researchers that are dominated by a limited number of individual authors, who are involved in majority of the group's publications. 
%The amount of collaborations in such environment is limited and the overlapping percentage among (co)authoring teams is very high. 
This is in contrast with decentralized communities where the authoring teams tend to be distinct or with small number of common individuals. Centrality limits the sources of knowledge generation in the field, lowering creativity and narrowing down the range of methodologies used in practice, which in turn could provoke the replicability crises~\cite{danchev2019centralized}. 

Since centrality is a localized property of scientific communities, we calculated this property at the subfield level rather than for entire main disciplines. We divided each of the 10 fields we used in this study into subfields using the most granular field tag assigned to papers by MAG (i.e., level 5). There exists a hierarchical structure to the field tags such that level-1 subfields (e.g., ``Quantum Mechanics'') have level-0 parents (e.g., ``Physics''), level-2 subfields (e.g., ``Hypergravity'') have level-1 parents (e.g., ``Quantum Mechanics''), and so on. We considered a level-5 tag $f_5$ to belong to a level-0 discipline $f_0$ only if $f_0$ was one of the ancestors of $f_5$. The centrality of a subfield community associated with a particular level-5 tag could be computed based on the coauthorship network of papers, where nodes represent contributing authors and the weight of an edge between two authors indicates the number of papers they co-authored (possibly with other authors). Networks that are closer to a star graph are more central than those with uniform edge distributions~\cite{kunegis2012fairness}. For each subfield, we obtained Lorenz curve of the coauthorship network and thereby computed the underlying Gini coefficient (see Appendix), a variable between 0 and 1 that indicates the network's centrality with 0 indicating no centrality and 1 representing maximum centrality (equivalently, star graph).

\subsubsection{Echo-Chamber Effect}
\label{subsubsec:echo_chamber}
Echo-chamber effect is the lack of exposure of a subfield's authors to ideas originating from nearby subfields or other disciplines. This occurs in scientific communities with minimal knowledge exchange with researchers that work on other topics~\cite{shi2019wisdom}. Echo-chamber effect confines collective creativity to the boundaries of the subfield's community and, similar to centrality, reduces the diversity of perspectives in the community~\cite{cinelli2021echo}. However, unlike centrality, it cannot be computed only through the within-community's coauthorship network and depends on cross-subfield collaborations, i.e., volume of coauthorship edges across subfields. Note that centrality and echo-chamber effects are mutually exclusive features of scientific communities and are not necessarily correlated (see Appendix~\ref{subapp:echo_chamber}). Inspired by how echo-chamber effect is measured for social media platforms~\cite{cota2019quantifying, gao2023echo}, we computed it as the volume of coauthorships between authors of a given community and those in other subfields divided by total number of coauthorships taken place in that community (see Appendix~\ref{subapp:echo_chamber}).
% groups created under the echo-chamber effect would be homogenous and highly polarized

\subsubsection{Social Media Posts}
\label{subsubsec:social_media_posts}
In this work, we used the number of times scientific articles are mentioned in tweets as indicator of attention they have received in social media. The count of mentions are extracted from SciSciNet database~\cite{lin2023sciscinet}. Based on the date of construction of this database, our data reflected only the tweets that had been posted by the year 2023.

\subsubsection{Other Metrics}
Other metrics including the number of authors, citation counts of papers and the prestige of publishing venues could be directly retrieved from MAG database. Specifically, we used the pre-computed rank scores for scientific journals included in MAG as a surrogate measure of their prestige perceived by scientific communities. This score computes the log-likelihood of the academic importance of the venues.

%Materials and Methods should be described with sufficient details to allow others to replicate and build on published results. Please note that publication of your manuscript implicates that you must make all materials, data, computer code, and protocols associated with the publication available to readers. Please disclose at the submission stage any restrictions on the availability of materials or information. New methods and protocols should be described in detail while well-established methods can be briefly described and appropriately cited. Research manuscripts reporting large datasets that are deposited in a publicly avail-able database should specify where the data have been deposited and provide the relevant accession numbers. If the accession numbers have not yet been obtained at the time of submission, please state that they will be provided during review. They must be provided prior to publication.

\begin{figure}[t]
\includegraphics[width=\columnwidth]{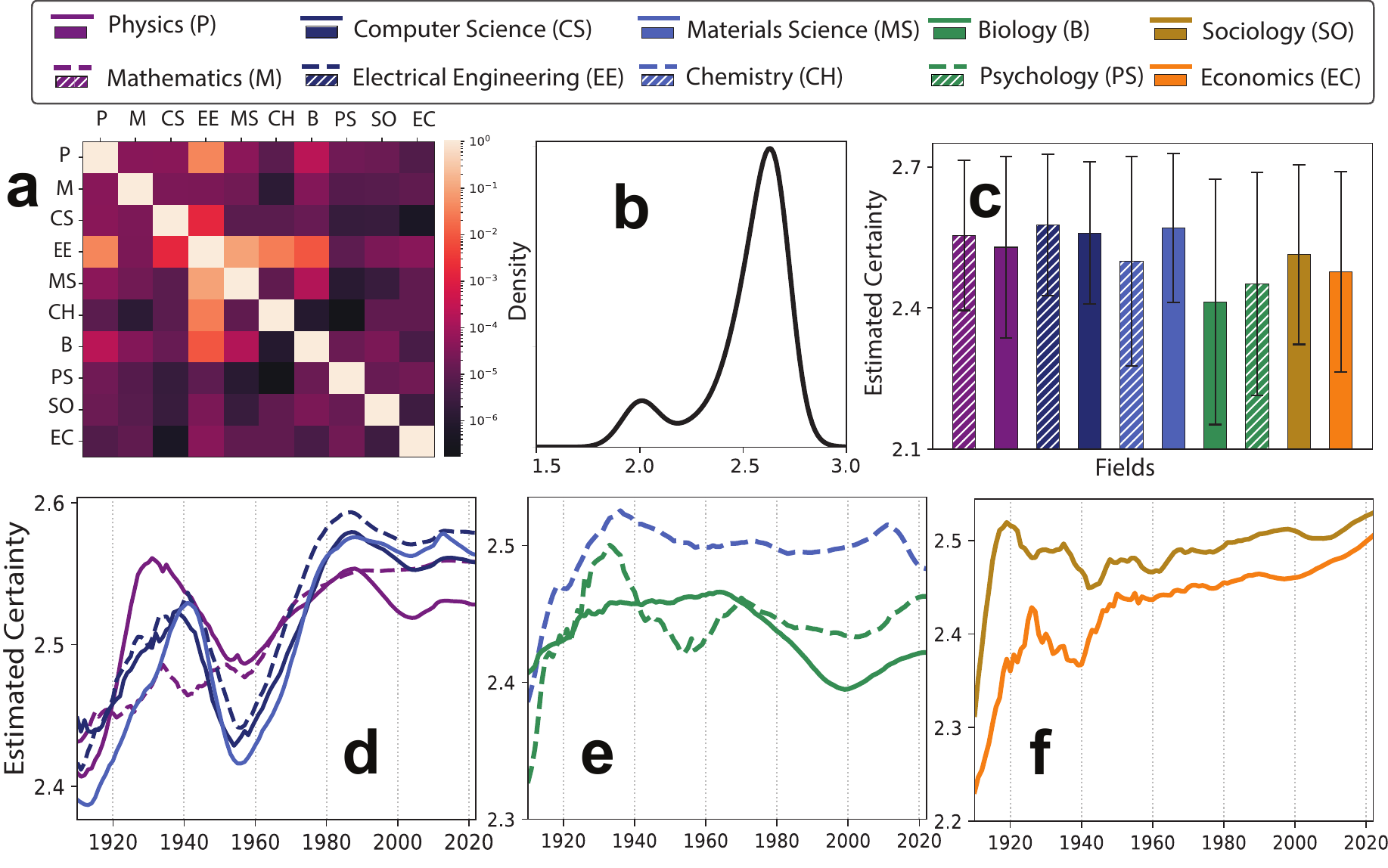}
\caption{\textbf{(a)} overlapping percentage between pairs of disciplines involved in our study. \textbf{(b--c)} Distribution of estimate language certainty: (\textbf{b}) the distribution of certainty measure for all papers involved in this study, and (\textbf{c}) average and standard deviation of measurements in each field. \textbf{(d--f)} temporal patterns of quantified certainty scores for computational fields \textbf{(d)}, life sciences \textbf{(e)} and social sciences\textbf{(f)}.
\label{fig:fieldbar_wholedist}}
\end{figure}

%%%%%%%%%%%%%%%%%%%%%%%%%%%%%%%%%%%%%%%%%%
\section{Results}

\subsection{Certainty Measurement}
Correlation coefficients between the certainty measurements and our annotations are shown in Figure~\ref{fig:methods_perf}(b) indicating that the two models SciBERT and LSTM significantly outperformed other methods, with the former being the overall winner with a slight margin. The distribution of estimated certainty scores obtained by all the competing models are also shown in Figure~\ref{fig:methods_perf}(b). Unlike other methods, SciBERT and LSTM led to bimodal distributions, where the higher peak in the former is located in less extreme values. Based on these results, we used SciBERT as the winning model for the rest of our analysis in this work.

\begin{figure}[t]
\begin{adjustwidth}{-\extralength}{0cm}
\centering
\includegraphics[width=1.15\columnwidth]{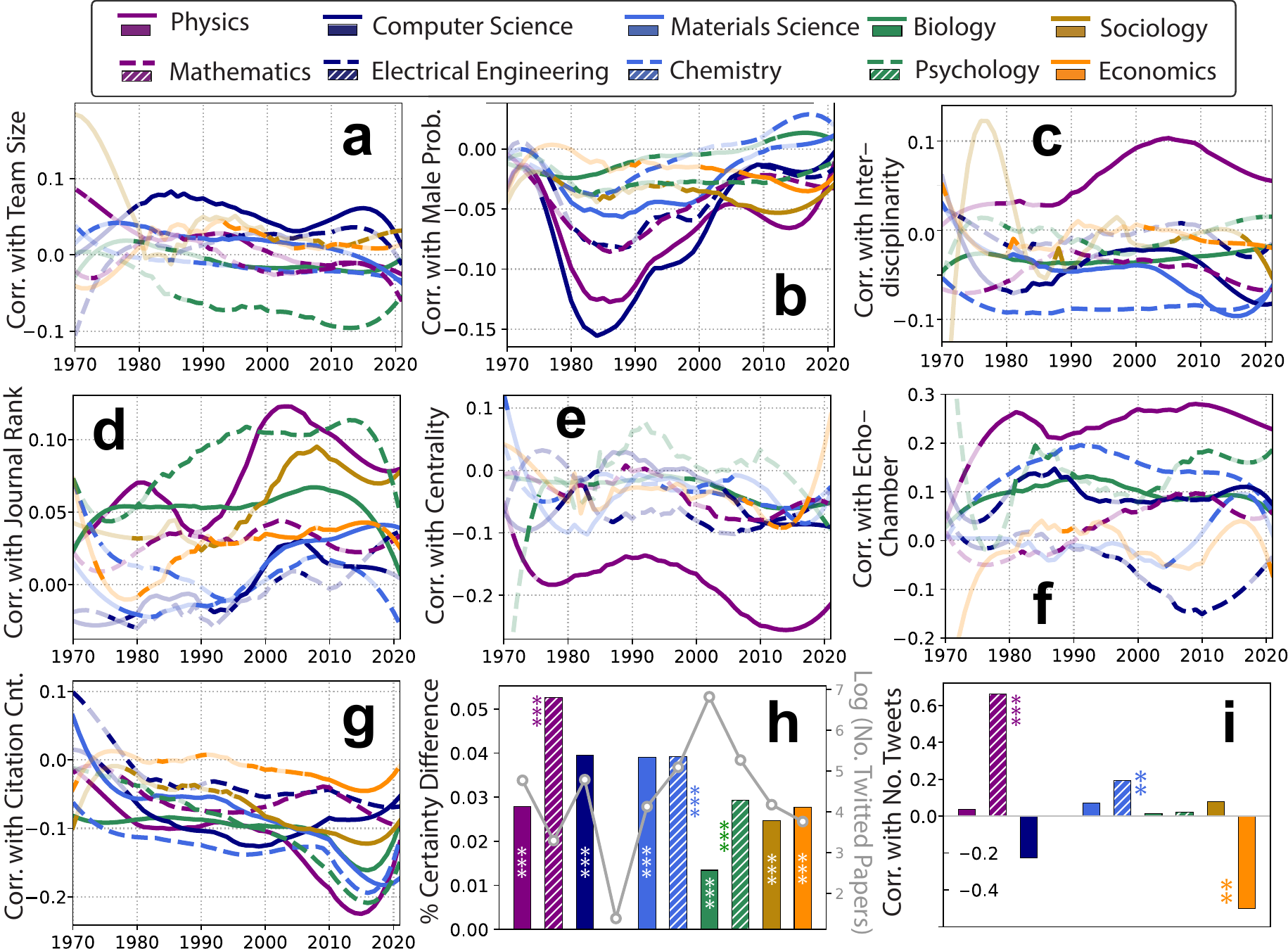}
\caption{Correlation analysis of language certainty with respect to bibliographic metrics: (\textbf{a--g}) temporal (partial) correlations between measured certainty and \textbf{(a)} number of authors (team size), \textbf{(b)} probability that the first author is male, \textbf{(c)} interdisciplinarity of authors team, \textbf{(d)} journal rank, \textbf{(e)} centrality of subfield's network, \textbf{(f)} echo-chamber effect, and \textbf{(g)} citation counts. (\textbf{h}) percentage of decrease in language certainty of papers with at least one twitter mention against those without any mentions (left axis), and logarithm of papers with at least one twitter mention per field (right axis). (\textbf{i}) partial correlations between language certainty and number of twitter mentions after controlling for the rank of publishing journal.
\label{fig:temporal_patterns}}
\end{adjustwidth}
\end{figure}

%\subsection{Field-wise, Non-temporal Analysis}
We compared the average SciBERT-measured certainty for all papers with at least one conclusive sentence in each of the 10 fields reported in Table~\ref{tab:fields}. The percentage of pairwise overlaps between these disciplines, measure in terms of Jaccard index, are shown in Figure~\ref{fig:fieldbar_wholedist}(a). The average Jaccard indices is very low--about 0.35\%. The field of electrical engineering had the highest Jaccard index with other fields including materials science, physics, chemistry, computer science and even plant biology, reflecting the fact that engineering methods are applied in a wide spectrum of disciplines. But even electrical engineering had only 3.5\% pairwise Jaccard index at its peak revealing a high level of distinctness among the chosen subfields. Applying SciBERT on this data resulted in certainty scores with a distribution shown in Figure~\ref{fig:fieldbar_wholedist}(b), which displays two peaks similar to but not as pronounced as SciBERT outputs on the validation data set (see Figure~\ref{fig:methods_perf}(b)). Figure~\ref{fig:fieldbar_wholedist}(c) shows the average and standard deviation of certainty scores across the selected fields. The results indicate that relatively more quantitative physical sciences (physics, mathematics, computer science, electrical engineering, materials science and chemistry) often verbalize their conclusion with higher certainty than non-quantitative physical sciences (i.e., biomedical sciences involved in our study: biology and psychology) and social sciences (sociology and economics). Moreover, on average, authors in the field of biology reported their findings with lowest subjective certainty among the 10 selected disciplines.

\begin{figure}[h]
\begin{adjustwidth}{-\extralength}{0cm}
\centering
\includegraphics[width=1.\columnwidth]{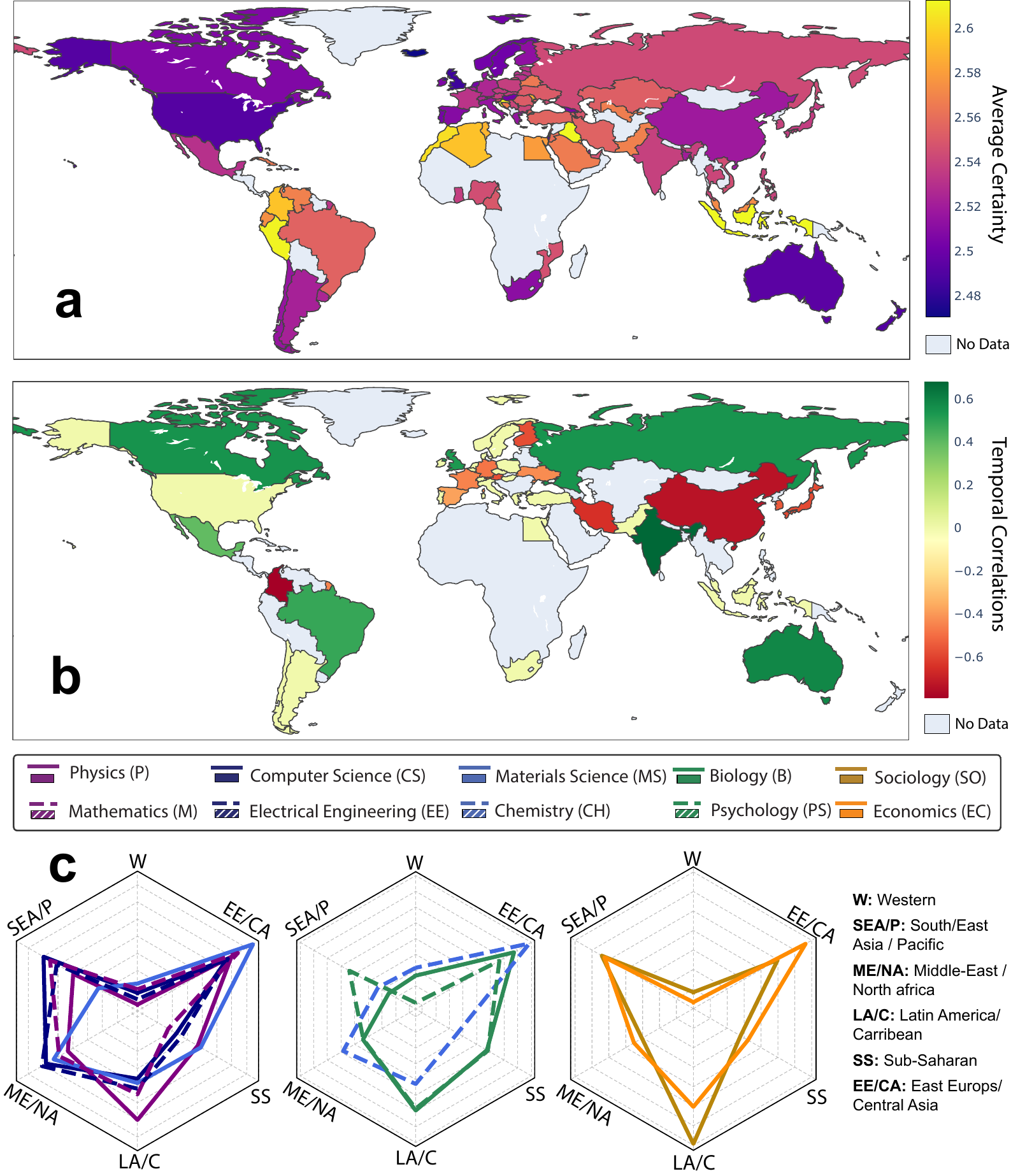}
\caption{\textbf{(a)} Worldwide heatmap of average subjective certainty in physics publications for countries that have been assigned at least 50 papers with certainty measurements. Darker colors represent lower average certainty. This heatmap considered all publications involved in our dataset disregarding the year of publication. \textbf{(b)} Spearman correlations between annual average certainty of publications and the year of publication, shown separately for each country. Correlations whose $p$-values were larger than 0.1 were excluded from the heatmap. \textbf{(c)} Certainty scores of articles averaged across six regional groups.
\label{fig:world_map}}
\end{adjustwidth}
\end{figure}

\subsection{Temporal Analysis}
We studied the temporal dynamics of subjective uncertainty in the selected 10 disciplines by computing the annual average of the calculated certainty scores. We then plotted these averages for 112 years, from 1910 to 2021. The second row in Figure~\ref{fig:fieldbar_wholedist} shows the resulting patterns separately for heavily computational sciences (Figure~\ref{fig:fieldbar_wholedist}(d)), less computational life sciences (Figure~\ref{fig:fieldbar_wholedist}(e)) and social sciences (Figure~\ref{fig:fieldbar_wholedist}(f)). There are two waves of increasing trends distinguishable in the certainty of computational sciences: from 1910 to around 1940, and from 1960 to 1980, after which the certainty scores became relatively flat. A sharp decrease in certainty scores was observed between 1940 and a year between 1955 and 1960 for most computational sciences (except math). This remarkable pattern first happened in physics starting in the late 1930's which was then followed by other technical fields such as electrical engineering and materials science. Investigating physics' subfields revealed that the topics that highly contributed to this decreasing pattern were mostly relevant to quantum (see Supplementary Materials). On the other hand, disciplines in life sciences and social sciences showed slower changes, except for the first 10 years between 1910 and 1920--this could be due to the artifact of having small sample size.

\subsection{Correlation Analysis}
Next, we analyze the association between certainty scores of abstracts to a set of bibliographic metrics. These associations are measured by means of Spearman correlation, except for cases when we control for other variables. In such cases, where association with respect to metrics are expected to have mutual influence, correlations were measured in a controlled manner through Pearson partial correlation~\cite{mohr2021statistical}. In this section, number of authors (team size) and interdisciplinarity of the authors team are taken as mutually influential variables, hence we measured correlation between certainty and interdisciplinarity when controlling for the team size and vice versa. We made the measurements and stored them separately for every year in the 50-year window $[1970, 2020]$. We did not consider years earlier than 1970 as we saw a sharp decay in the number of annual publications in our database which made our correlation analysis statistically insignificant. In the temporal correlation curves shown in Figure~\ref{fig:temporal_patterns}, we blurred out correlations for years for which the corresponding $p$-value was not significant (using level $\alpha=0.05$). 

For pre-publication metrics, we formed the coauthorship network based on publications between years $t-10$ and $t-1$ in the subfield, and computed their correlations with certainty scores calculated for publications in year $t$. This process gave us one correlation per year $t$, and repeating it for different years $t$ led to our temporal correlations curve. We ignored periods with sparse coauthorship graphs that had fewer than 50 author nodes.

\textit{\textbf{Publication Metrics--}}Year-wise correlations between certainty and metadata of the publications are shown in Figure~\ref{fig:temporal_patterns}(a--d). One can observe that journal rank and male probability showed significant correlations for more disciplines. Here, we considered male probability of the first authors, however, we found similar temporal correlation patterns when computing male probability of the last and all authors (see Supplementary Materials). Figure \ref{fig:temporal_patterns}(b) shows that the correlations are highly negative for computational sciences between 1980 and 2000, and the magnitude of correlations decayed for the last 20+ years. Moreover, highly positive correlations with respect to journal rank could be observed for social sciences. Biology and physics also resulted in significant correlations as well, with the former showing a sharp decay in the last 10 years. For the other two publication metrics, most correlations were small in magnitude, however, there were a few exceptions in both. For example, while interdisciplinarity showed slightly negative correlations for most fields, Physics showed a considerable positive correlations over most years.

\textit{\textbf{Pre-publication Metrics--}}Correlations between subjective language certainty and pre-publication metrics are shown in Figure~\ref{fig:temporal_patterns}(e--d). The annual correlations with respect to the subfield's centrality are insignificant for most fields except for physics, which showed highly negative correlations. On the other hand, the annual correlations of the echo-chamber effect are significantly positive for most STEM fields, except for electrical engineering. Moreover, there was no consistent temporal trend over the past 50 years. Hence, in most cases, the stronger the echo-chamber effect in a subfield, the greater the subjective certainty reflected in its scientific language. 

\textit{\textbf{Post-publication Metrics--}}Temporal correlation patterns between subjective certainty and citation counts in Figure~\ref{fig:temporal_patterns}(g), showing a clear decreasing correlations into significantly negative values. For electrical engineering and materials science, this correlation used to be significantly positive 50 years ago. But by the mid-2010's, all disciplines showed significantly negative correlations, meaning that publications with lower language certainty are likelier to be cited in most fields. However, the magnitude of these correlations has been decreasing over the last 5-7 years. We also investigated associations between language certainty and the amount of social media attentions. Our investigation here was not temporal and focused on the year 2017, mainly because we found very few papers of our study mentioned in at least one tweet except for year 2017. %Total number of papers with corresponding tweets is displayed in Figure~\ref{fig:temporal_patterns}(h) in logarithmic scale (right axis). 

First, we categorized the papers into two groups: those that were not mentioned in any tweets and those with at least one tweet. Comparing these two groups revealed that the latter had significantly lower subjective certainty than the former based on Mann-Whitney U test (Figure~\ref{fig:temporal_patterns}(h), left axis). %Note that in this result, electrical engineering is absent due to having very few papers in SciSciNet with at least one tweet. 
Having observed this difference, we also sought for any associations between the number of tweets and language certainty. Figure~\ref{fig:temporal_patterns}(i) shows the field-wise partial correlations between language certainty and number of tweets after controlling for the rank of publishing journals, as we readily expect the quality of journal to positively influence the volume of external attention. We can distinguish two opposing trends: mathematics and chemistry showed significantly positive correlations, hence higher language certainty in these fields is likely to receive higher twitter visibility, while in economics, higher language certainty attracted less social media attention. Other fields did not show any significant correlations, though there existed a considerable negative correlation for computer science.

\subsection{Geographical Analysis}
\label{subsec:geography}
Geographical distribution of average language certainty of articles published in physics is shown in Figure~\ref{fig:world_map}(a). We ignored countries with less than 50 publications in this visualization. The result indicates that papers published in Western countries, including western Europe, North America and Oceania, generally exhibited lower certainty than the majority of other countries. However, China shows a similarly low level subjective certainty. Obtaining the worldwide certainty heatmap for the other nine disciplines demonstrated similar patterns with the exception of computational fields like computer science and electrical engineering, which exhibited smaller disparity between developed and developing countries (see Supplementary Materials). 

In order to summarize the geographical distribution of subjective certainty of papers published in various fields, we divided the countries into six groups and took the total average of certainty scores measured for institutions across these geographical regions~\cite{unicef} (Supplementary Materials). Figure~\ref{fig:world_map}(c) shows the normalized region-wise averages separately for computational sciences (left), life sciences (center) and social science (right). Clearly, Western countries consistently showed the lowest average subjective certainty, while Eastern Europe/Central Asia (EE/CA) resulted in the highest level of average certainty across all disciplines. The discrepancies are slightly less pronounced for publications in life sciences.

We also examined the temporal dynamic of average certainty of physics articles written in different countries in the span of 20 years between years 2000 and 2020. We presented this result in another worldwide heatmap in Figure~\ref{fig:world_map}(b), where the colors represent Spearman correlation between the annual average certainty of each country and the years. We disregarded correlations with insignificant $p$-values ($>0.1$). This result indicates a more moderate separation between Western and non-Western countries. For example, whereas the United States had a flat dynamic and the United Kingdom exhibited growth in certainty, many other European countries showed a decreasing trend in the considered 20-year window. %The three countries with the \emph{most rapid decay} in language certainty in physics were Colombia, China and Iran, and those with the \emph{mast rapid growth} in certainty were India, Australia and Canada.

%%%%%%%%%%%%%%%%%%%%%%%%%%%%%%%%%%%%%%%%%%
\section{Discussion}

The results obtained primarily showed that the quantified subjective certainty scores of publications in computational sciences---including physics, mathematics, computer science, electrical engineering, and materials science---are, on average, higher than other disciplines in our study. The temporal patterns of language certainty in these computational fields were mostly increasing except for two time windows. The more recent one started about 40 years ago and continued until the last year in our dataset. In this period subjective certainty of these fields became relatively flat. The older window began almost simultaneously with World War II, marking a rapid decline in subjective certainty of physics publications. Since quantum-related subfields contributed the most to the rise of this wave of uncertainty in physics, one may attribute this to the side-effect of World War I on German scientists and their contribution to quantum mechanics~\cite{forman2013weimar}. This provides a more compelling explanation than directly relating these dramatic declines to World War II, as it can also explain the delayed emergence of a similar trend in other computational fields. On the other hand, less computational fields in our study showed more stable dynamics, except for a substantial certainty decrease in biology publications after 1970, and a slow but steady certainty growth in social sciences after 1980.

Our correlation analysis of pre-publication metrics revealed that, as expected~\cite{difonzo2013rumor}, having echo chambers effects within a subfield's community can boost the subjective certainty of researchers. Regarding publication metrics, we observed that female scientists are more likely to use higher subjective certainty in their academic language than male researchers. This discrepancy was highest between 1980 and 1990, especially in computational fields like computer science and physics, but it subsided a lower level by early 2000's.  Moreover, journals in physics as well as less computational fields, such as psychology and sociology, were slightly inclined towards publications with higher language certainty. Furthermore, studying post-publication metrics indicated that articles with lower subjective certainty are more likely to receive attention from academic and/or lay audience.

In our final set of results, plotting the geographical distributions showed a substantial disparity in the level of quantified certainty of articles published in different countries. Specifically, we observed that, on average, publications in which the first and last authors are affiliated with institutions in western countries exhibited lower certainty. However, there is no consistent temporal dynamic in annual averages of certainty scores across countries.

We admit that our certainty quantification method is not perfect and can be improved. At least two directions could be followed to enhance the quality of the certainty estimates: (1) improving the language model measuring the underlying subjective (un)certainty, and (2) extending the application of such methods to full texts of the articles rather than the abstracts. Furthermore, it would be insightful to study potential correlations between subjective and objective uncertainty of scientific publications.

%%%%%%%%%%%%%%%%%%%%%%%%%%%%%%%%%%%%%%%%%%
%\section{Conclusions}

%This section is not mandatory, but can be added to the manuscript if the discussion is unusually long or complex.

%%%%%%%%%%%%%%%%%%%%%%%%%%%%%%%%%%%%%%%%%%
\vspace{6pt} 

\appendixtitles{yes} % Leave argument "no" if all appendix headings stay EMPTY (then no dot is printed after "Appendix A"). If the appendix sections contain a heading then change the argument to "yes".
\appendixstart
\appendix
\section[\appendixname~\thesection]{Uncertainty Quantification}
\label{app:models}
\subsection[\appendixname~\thesubsection]{Finetuned SciBERT}
We used a SciBERT model fine-tuned by~\citet{pei2021measuring} based on their own uncertainty annotations over sentences chosen from scientific publications and news outlets. Their training data were prepared based on six-level annotations, hence the output of the model is a scalar between 0 and 6. Here, we used their sentence-based uncertainty estimator.

\subsection[\appendixname~\thesubsection]{Zero-Shot Classification}
Our zero-shot method was based on zero-shot classification of sentences into two categories: ``certain'' or ``uncertain'' sentences~\cite{yin2019benchmarking}. For this purpose, we used a BART-large model in the context of multi-genre natural language inference (MNLI) to infer entailment relationship between the hypothesis , i.e., the sentence for which we are measuring uncertainty, and the two premises that are set to ``It is certain.'' and ``It is uncertain.''. The output of our zero-shot method would be the probability of entailment with regards to the first premise, hence between zero and one.

\subsection[\appendixname~\thesubsection]{Pre-trained LSTM}
We used an LSTM model with five layers that is pre-trained over a 3-class classification training data set, where each class represents a specific level of uncertainty of a sentence. The training data set included scientific sentences in biomedical disciplines. This model outputs three class probabilities for every input string. The final scalar uncertainty score is obtained by taking the expected value of this 3-class probability distribution.

\subsection[\appendixname~\thesubsection]{Generative AI}
We used the online service ChatGPT in September of 2023 to inquire about the level of language uncertainty of our annotated sentences from GPT-4 model. We used the following prompt to instruct the generative AI engine what data and in what format we expect:

\begin{quote}
    \textit{Research article abstracts summarize the findings of a scientific paper with language that expresses more or less certainty in the results and their implications for our understanding of the world. Evaluate the following sentences by describing in a short paragraph the level of certainty or uncertainty that their authors express about their conclusions, including words and phrases that lead you to the assessments. Then rate each sentence from 1 to 100 in terms of their certainty, where 1 expresses extreme uncertainty about the research findings and their implications, and 100 expresses extreme certainty in research results and their significance.}
\end{quote}
This introductory prompt would then be followed by the sentences in our evaluation dataset.

\section[\appendixname~\thesection]{Bibliographic Metrics}
\label{app:bib_metrics}

\subsection[\appendixname~\thesubsection]{Gender Inference}
\label{subapp:gender_infer}
We created the feature vectors by computing tf-idf (term frequency--inverse document frequency) metric for character $n$-grams, where we varied $n$ in the range 1 to 9. Each feature would then be scaled using their log-odds to improve performance of a logistic regression classifier~\cite{wang2012baselines, hu2021s}, trained by setting the objective to the $L_2$ penalized log-likelihood with hyperparameter $C=2$. The training is performed on 80\% of SSA database and tested on the remaining data, resulting in test $F_1$-score of 76\% (when threshold$=0.5$) and area under the curve (test AUC) of 81\%.

\subsection[\appendixname~\thesubsection]{Network Centrality}
\label{subapp:centrality}
Centrality of a graph measures the non-uniformity of distribution of edges over the nodes. A fully centralized network is a star graph with $v$ vertices, where all edges connect a central node to other $n-1$ vertices. In such network, the degree of the central node is $v-1$ and other nodes have degree 1, resulting in an \emph{unfair} edge distribution. On the other hand, least centralized graphs have close-to-uniform distribution of edges, i.e., node degrees are almost the same throughout the graph. The aim of centrality metrics are to quantify the proximity of a given graph to a star-like network with maximum centrality. In this work, we used Lorenz curves to quantify the uniformity of edge distribution. This curve was traditionally used in economic applications to measure the inequality of the distribution of wealth~\cite{sitthiyot2021simple}. In the context of network connectivities, the wealth is translated to the number of edges connected to each node. 

\begin{figure}[h]
\centering
\includegraphics[width=.9\columnwidth]{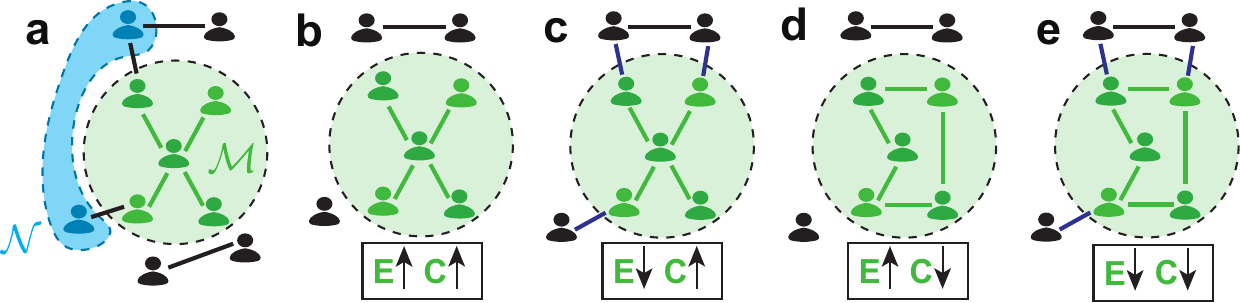}
\caption{Centrality (C) and echo-chamber effect (E) in different networks. \textbf{(a)} In each of the sample networks, like the one shown in (a), part of the network that is in green shaded region represents the community of the target subfield with the green nodes being part of the community (denoted by $\mathcal{M}$) and non-green nodes being part of a neighboring external community. The blue nodes are the neighbors that we would denote by $\mathcal{N}$ in the equation of echo chamber effect (see \ref{subapp:echo_chamber}). \textbf{(b)} both metrics are high; \textbf{(c)} centrality is high but echo-chamber effect is low; \textbf{(d)} centrality is low but echo-chamber effect is high; \textbf{(e)} both metrics are low.
\label{figapp:network_metrics}}
\end{figure}

The Lorenz curve could be obtained by first ordering the nodes from smallest to largest degrees, and then assigning the sorted nodes their cumulative normalized degrees. Connecting these points in a space where the $x$-axis is the index of sorted nodes, and the $y$-axis represents the cumulative normalized degrees leads to a convex curve starting from $(0,0)$ ending at $(1,1)$. In one extreme, Lorenz curve of a graph with all nodes having with the same degrees (hence, uniform edge distribution) is the diagonal line $y=x$, and on the other extreme, Lorenz curve of a star-network is a piecewise linear curve connecting three points $(0,0)$, $(1,0)$ and $(1,1)$. A popular metric for measuring the proximity of networks to star graph is the area between the diagonal line $y=x$ and the obtained Lorenz curve. This metric is called the \emph{Gini coefficient} and is between 0 (uniform distribution of edges) and 1 (star-graph with maximum centrality). In this work, we computed Gini coefficients on coauthorship networks, where each node represents an author and having an integer weight $w$ for an edge between two authors indicates that they coauthored $w$ papers.

\subsection[\appendixname~\thesubsection]{Network Echo-Chamber Effect}
\label{subapp:echo_chamber}
Echo-chamber effect measures how isolated a given community is with regards in the universal coauthorship network of a given science subfield. In this context, isolation means having no or very few connections with authors of other subfields. Let $\mathcal{M}$ denote the set of nodes associated with authors in our target community and $\mathcal{N}$ denote the set of \emph{external} nodes that are neighbors of at least one node in $\mathcal{M}$. In Figure~\ref{figapp:network_metrics}(a), authors in the community's network ($\mathcal{M}$) are shown in green and their neighbors ($\mathcal{N}$) are in blue. Then, the echo-chamber effect of $\mathcal{M}$ is defined as the ratio $\frac{|\mathcal{M}|}{|\mathcal{M}\cup\mathcal{N}|}$. For the example shown in Figure~\ref{figapp:network_metrics}(a), the echo-chamber metric is calculated as $\frac{5}{5+2}\approx0.714$. This definition implies that echo-chamber effect is high when the volume of connections between internal and external nodes of a community is significantly high in comparison to the size of the subfield's community.

Note that network centrality and echo-chamber effect are not necessarily correlated with each other. This is illustrated in Figure~\ref{figapp:network_metrics}(b–e), where all combinations of low and high levels of the two metrics are shown.

%The appendix is an optional section that can contain details and data supplemental to the main text---for example, explanations of experimental details that would disrupt the flow of the main text but nonetheless remain crucial to understanding and reproducing the research shown; figures of replicates for experiments of which representative data are shown in the main text can be added here if brief, or as Supplementary Data. Mathematical proofs of results not central to the paper can be added as an appendix.

%All appendix sections must be cited in the main text. In the appendices, Figures, Tables, etc. should be labeled, starting with ``A''---e.g., Figure A1, Figure A2, etc.

%%%%%%%%%%%%%%%%%%%%%%%%%%%%%%%%%%%%%%%%%%
\begin{adjustwidth}{-\extralength}{0cm}

\reftitle{References}


\bibliography{refs}

% If authors have biography, please use the format below
%\section*{Short Biography of Authors}
%\bio
%{\raisebox{-0.35cm}{\includegraphics[width=3.5cm,height=5.3cm,clip,keepaspectratio]{Definitions/author1.pdf}}}
%{\textbf{Firstname Lastname} Biography of first author}
%
%\bio
%{\raisebox{-0.35cm}{\includegraphics[width=3.5cm,height=5.3cm,clip,keepaspectratio]{Definitions/author2.jpg}}}
%{\textbf{Firstname Lastname} Biography of second author}

% For the MDPI journals use author-date citation, please follow the formatting guidelines on http://www.mdpi.com/authors/references
% To cite two works by the same author: \citeauthor{ref-journal-1a} (\citeyear{ref-journal-1a}, \citeyear{ref-journal-1b}). This produces: Whittaker (1967, 1975)
% To cite two works by the same author with specific pages: \citeauthor{ref-journal-3a} (\citeyear{ref-journal-3a}, p. 328; \citeyear{ref-journal-3b}, p.475). This produces: Wong (1999, p. 328; 2000, p. 475)

%%%%%%%%%%%%%%%%%%%%%%%%%%%%%%%%%%%%%%%%%%
%% for journal Sci
%\reviewreports{\\
%Reviewer 1 comments and authors’ response\\
%Reviewer 2 comments and authors’ response\\
%Reviewer 3 comments and authors’ response
%}
%%%%%%%%%%%%%%%%%%%%%%%%%%%%%%%%%%%%%%%%%%
% I commented out
%\PublishersNote{}
\end{adjustwidth}

\newpage
\begin{adjustwidth}{-\extralength}{0cm}
\section*{Supplementary Materials}
\subsection*{Annual Averages of Certainty for Subfields of Physics}
\begin{figure}[!h]
\begin{adjustwidth}{-\extralength}{0cm}
\centering
\includegraphics[width=.85\columnwidth]{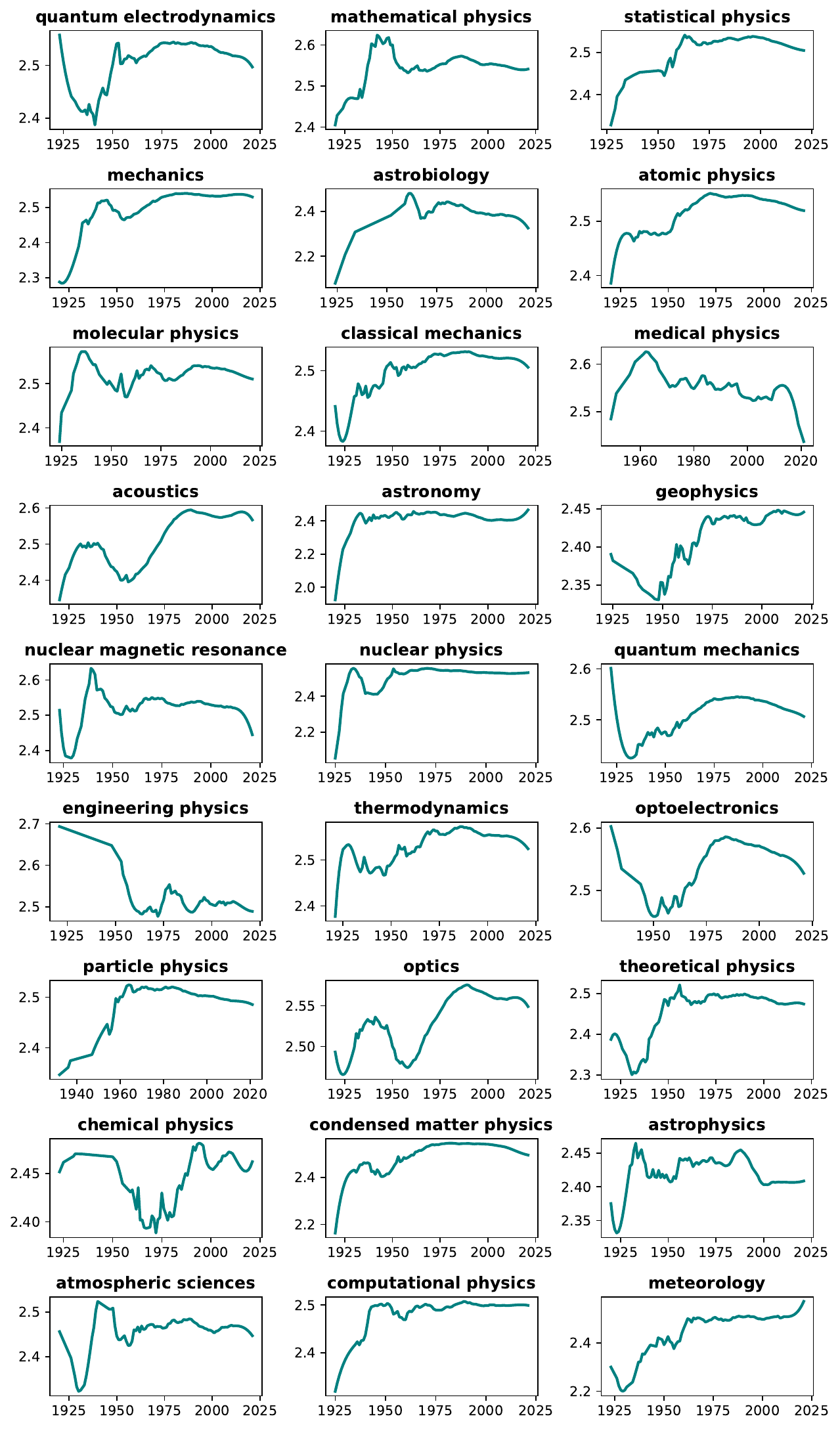}
\end{adjustwidth}
\end{figure}

\newpage
\subsection*{Temporal Correlations With Respect to Different Types of Estimates for Male Probabilities (see section 2.3.1)}
\begin{figure}[!h]
\begin{adjustwidth}{-\extralength}{0cm}
\centering
\includegraphics[width=1.05\columnwidth]{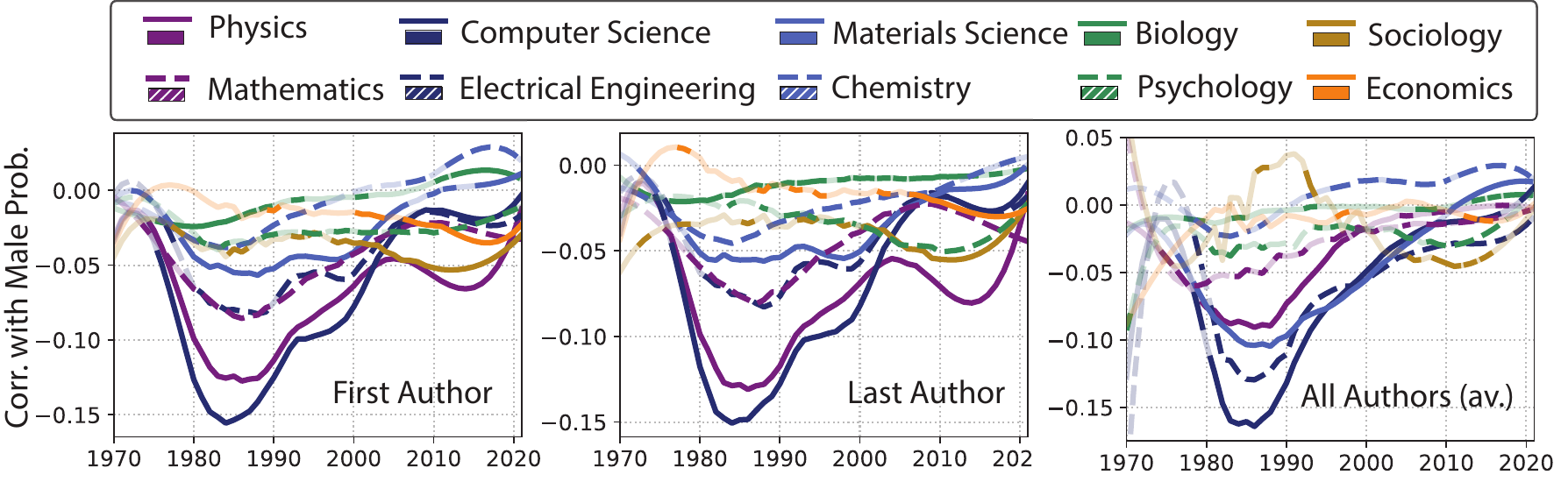}
\end{adjustwidth}
\end{figure}

\subsection*{Division of Countries into Six Regional Groups}
\begin{figure}[!h]
\begin{adjustwidth}{-\extralength}{0cm}
\centering
\includegraphics[width=.95\columnwidth]{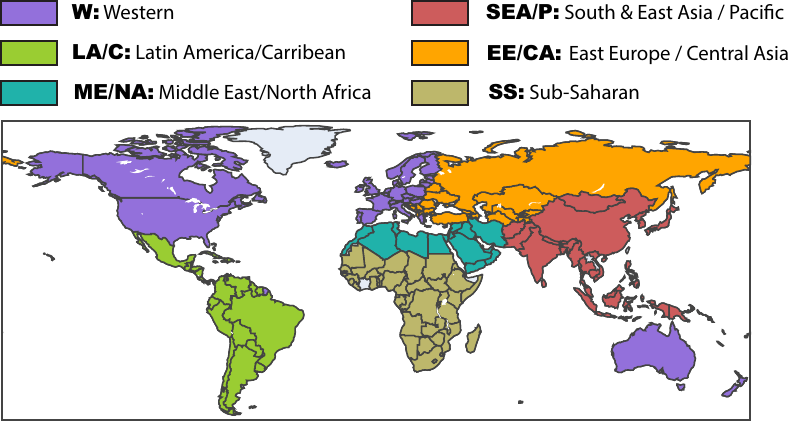}
\end{adjustwidth}
\end{figure}

\subsection*{Worldwide Distribution of Average Certainty for Disciplines Other than Physics}
\begin{figure}[!h]
\begin{adjustwidth}{-\extralength}{0cm}
\centering
\includegraphics[width=.95\columnwidth]{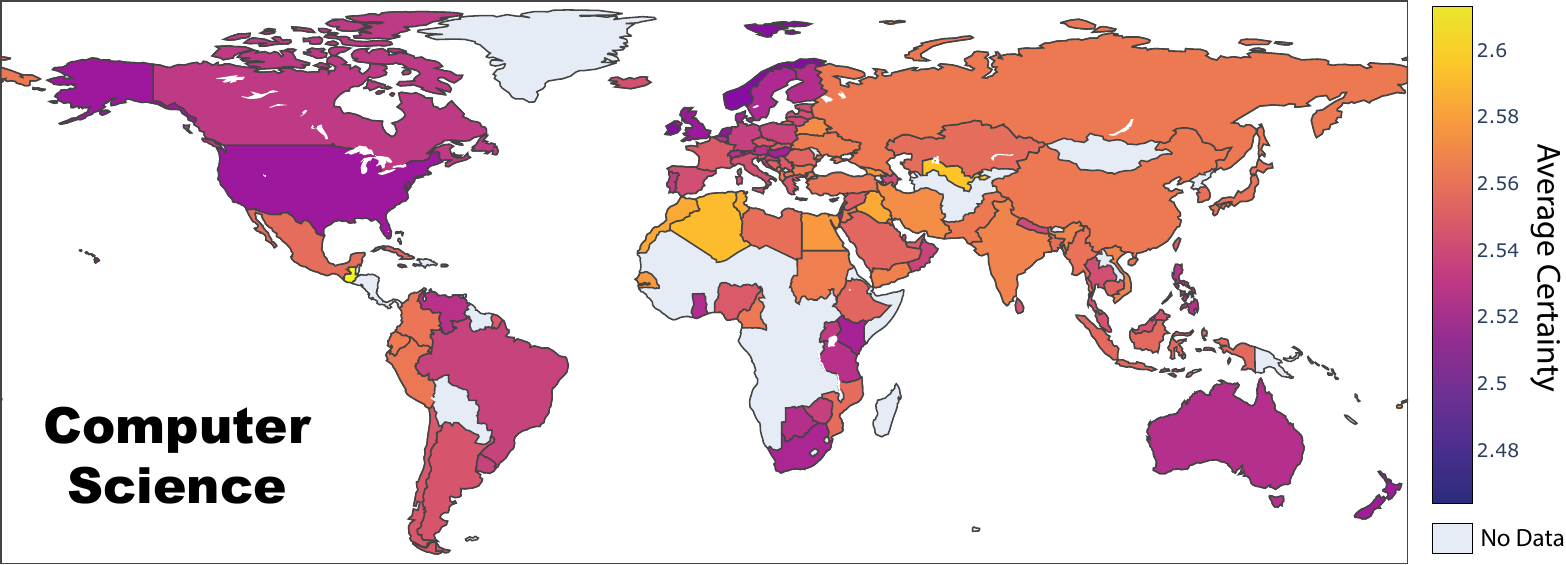}
\end{adjustwidth}
\end{figure}
\begin{figure}[!h]
\begin{adjustwidth}{-\extralength}{0cm}
\centering
\includegraphics[width=.95\columnwidth]{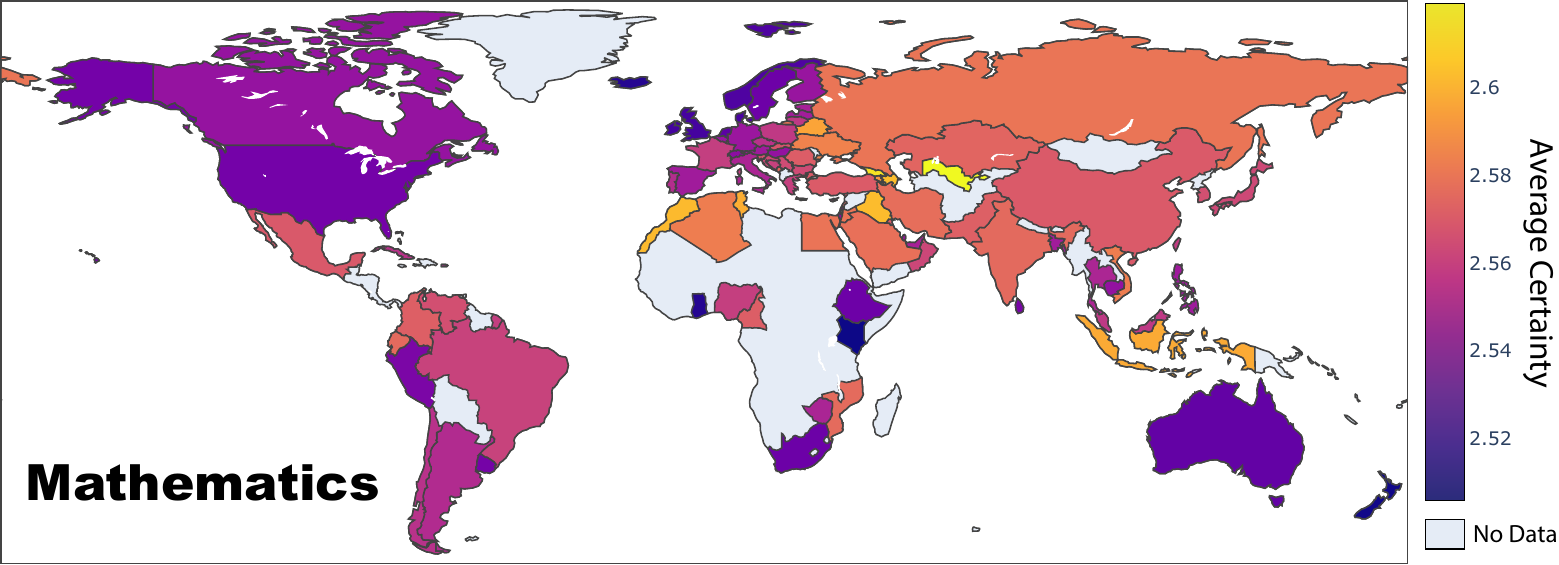}
\includegraphics[width=.95\columnwidth]{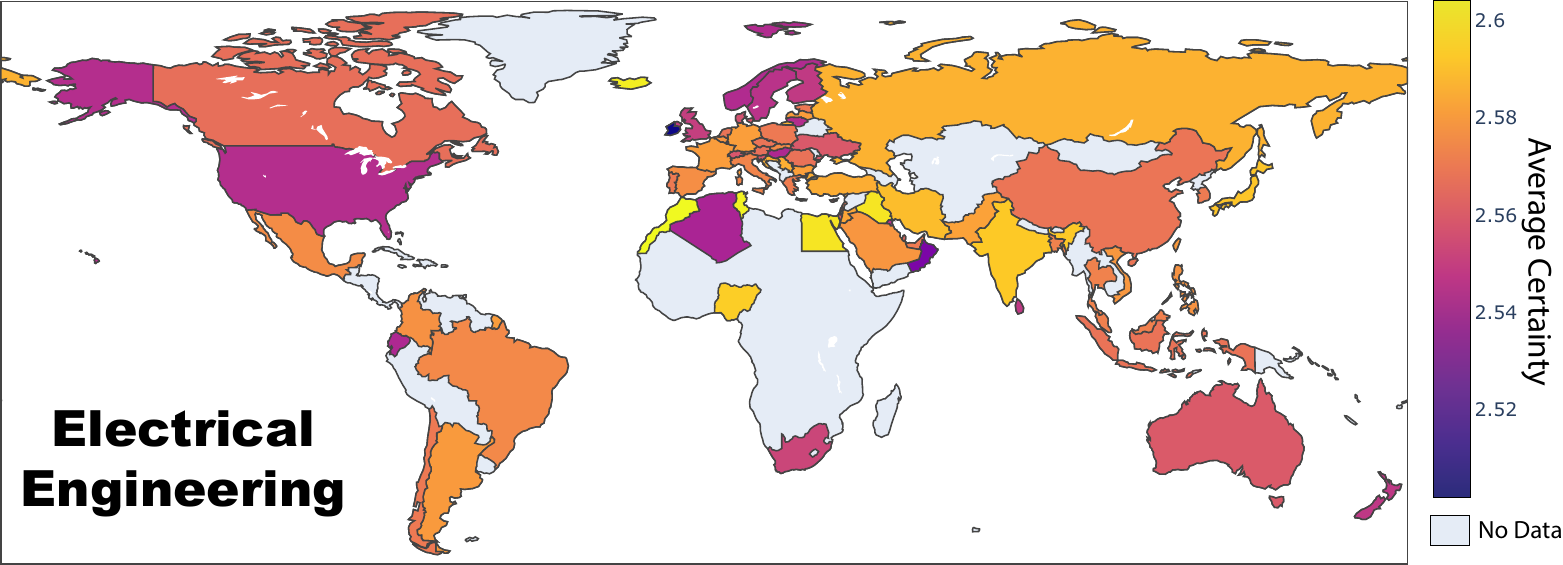}
\includegraphics[width=.95\columnwidth]{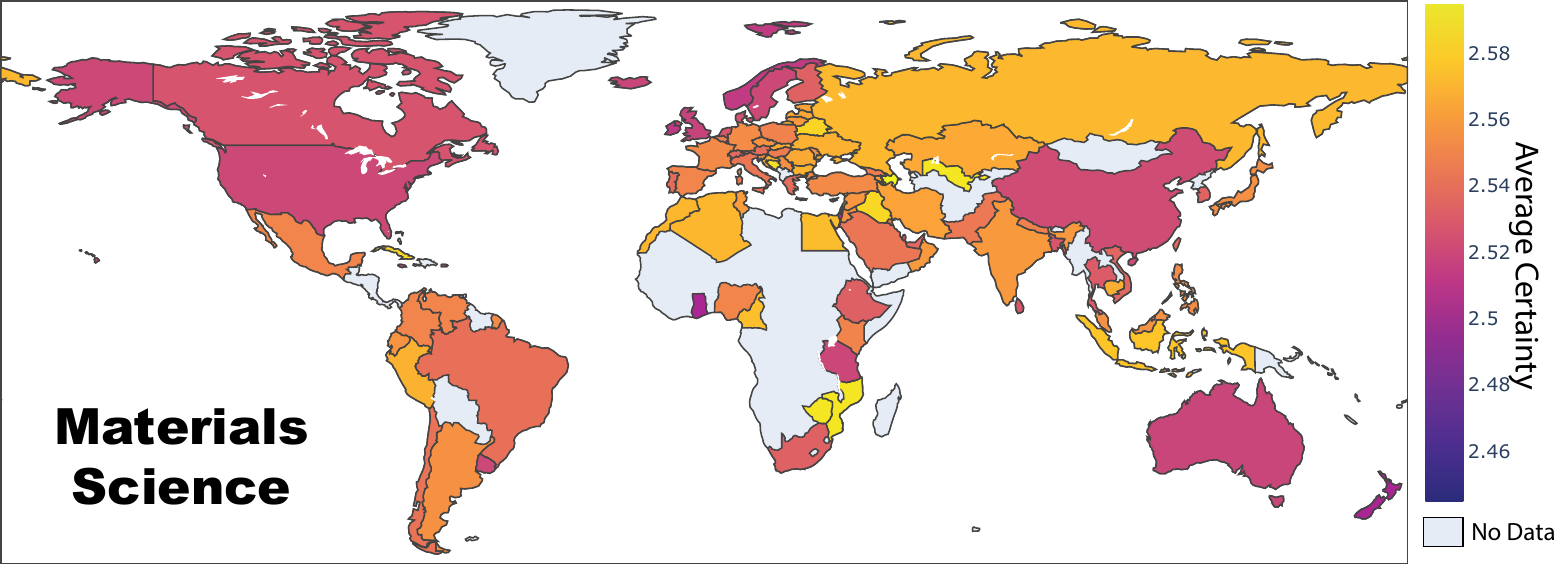}
\includegraphics[width=.95\columnwidth]{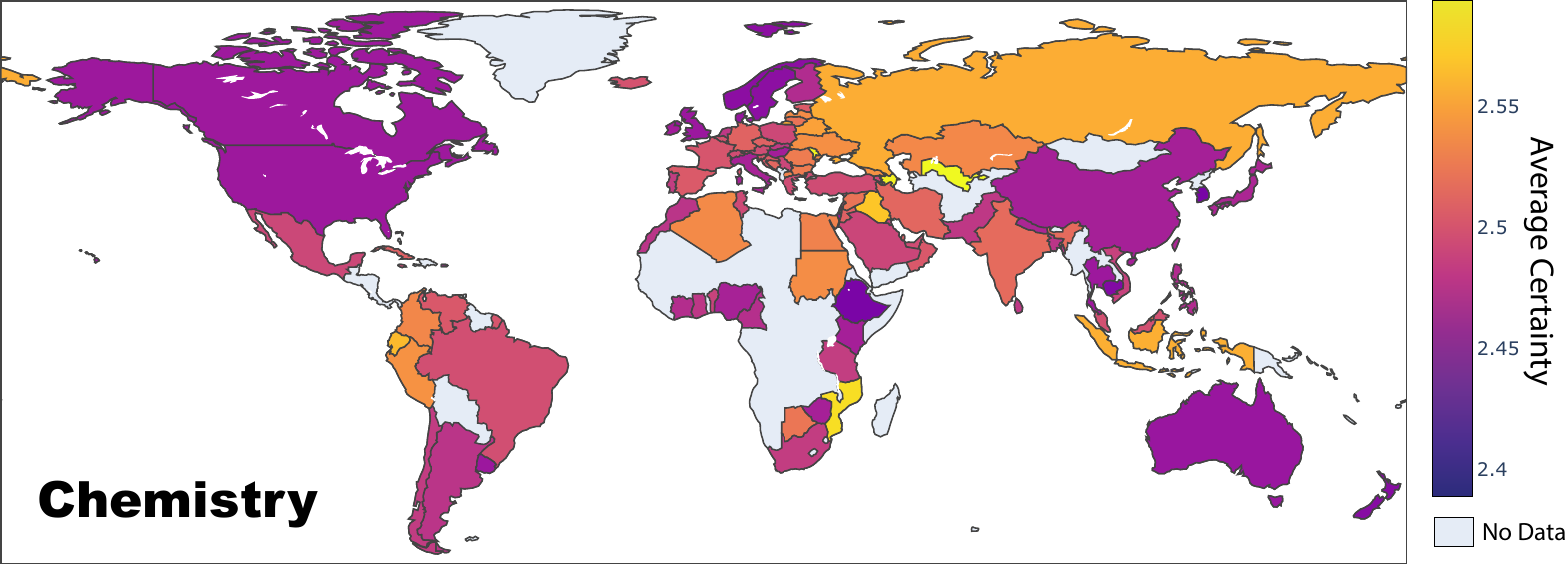}
\includegraphics[width=.95\columnwidth]{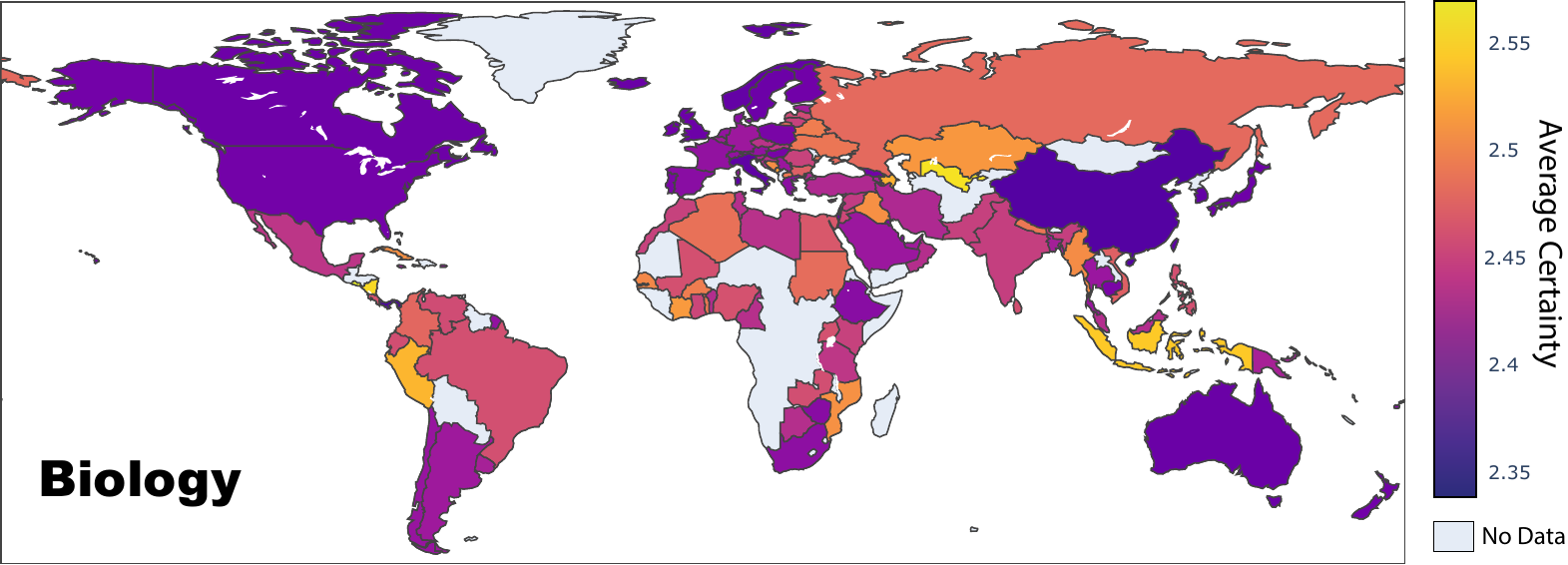}
\end{adjustwidth}
\end{figure}
\begin{figure}[!t]
\begin{adjustwidth}{-\extralength}{0cm}
\centering
\includegraphics[width=.95\columnwidth]{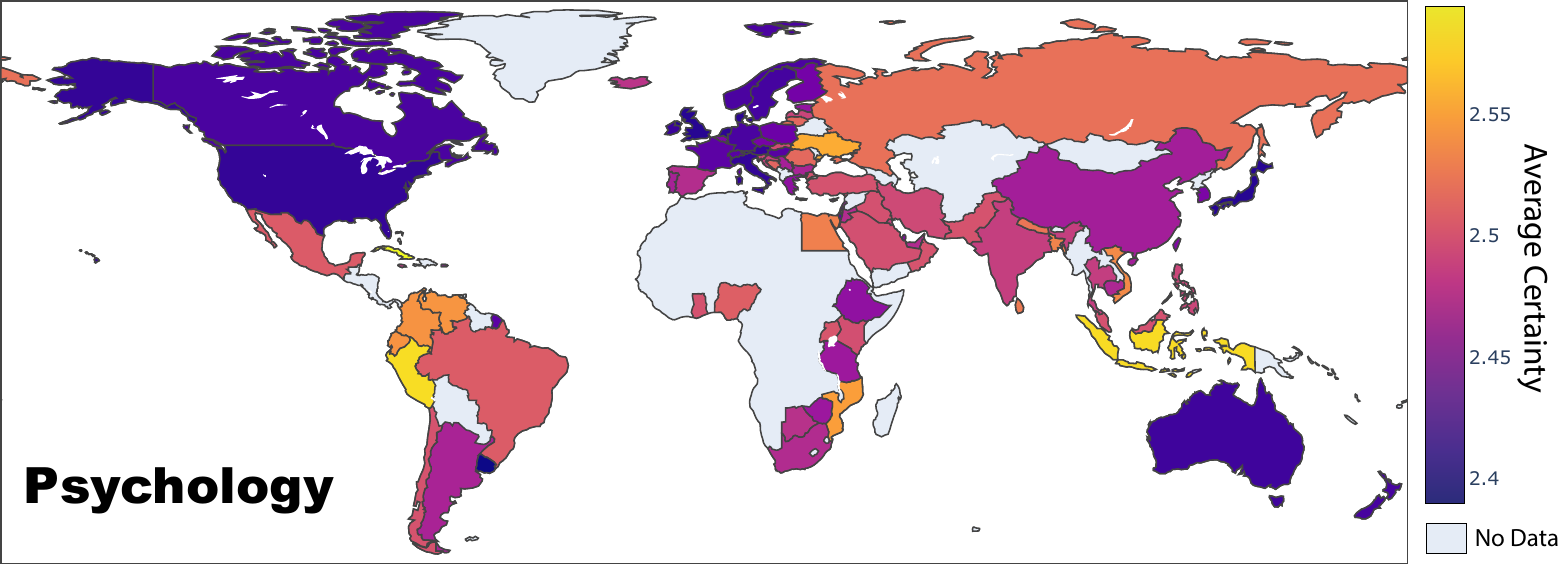}
\includegraphics[width=.95\columnwidth]{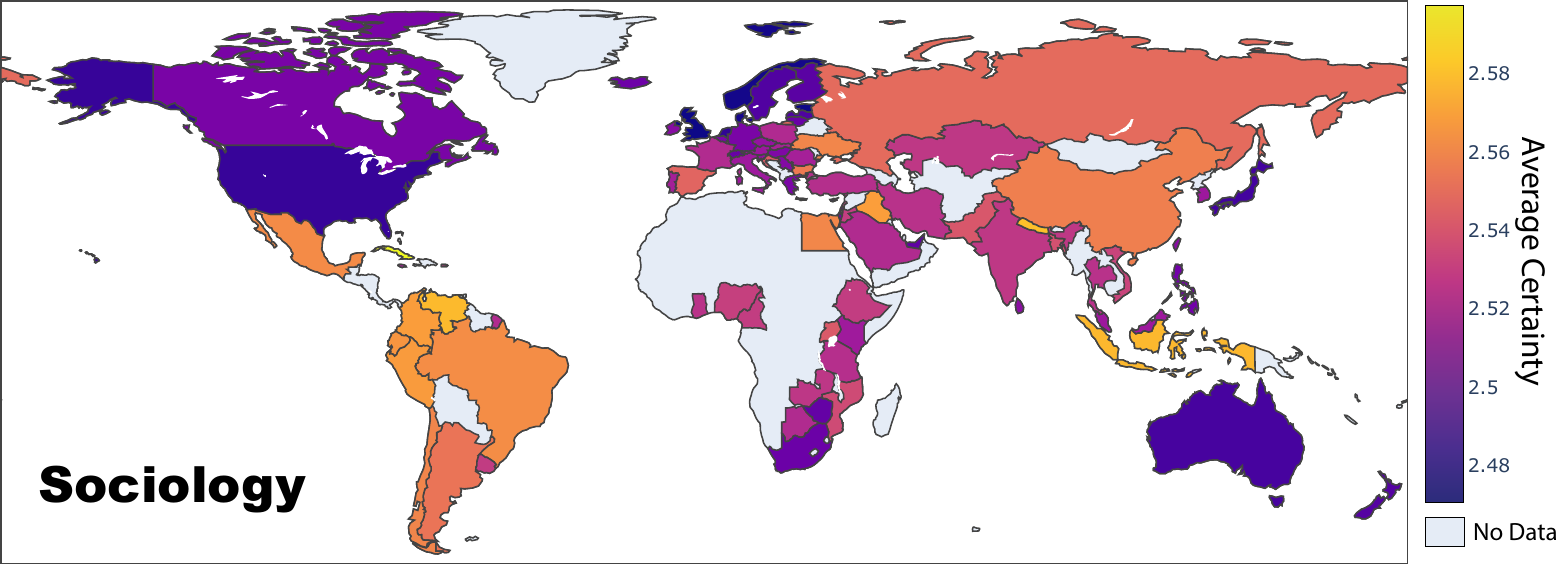}
\includegraphics[width=.95\columnwidth]{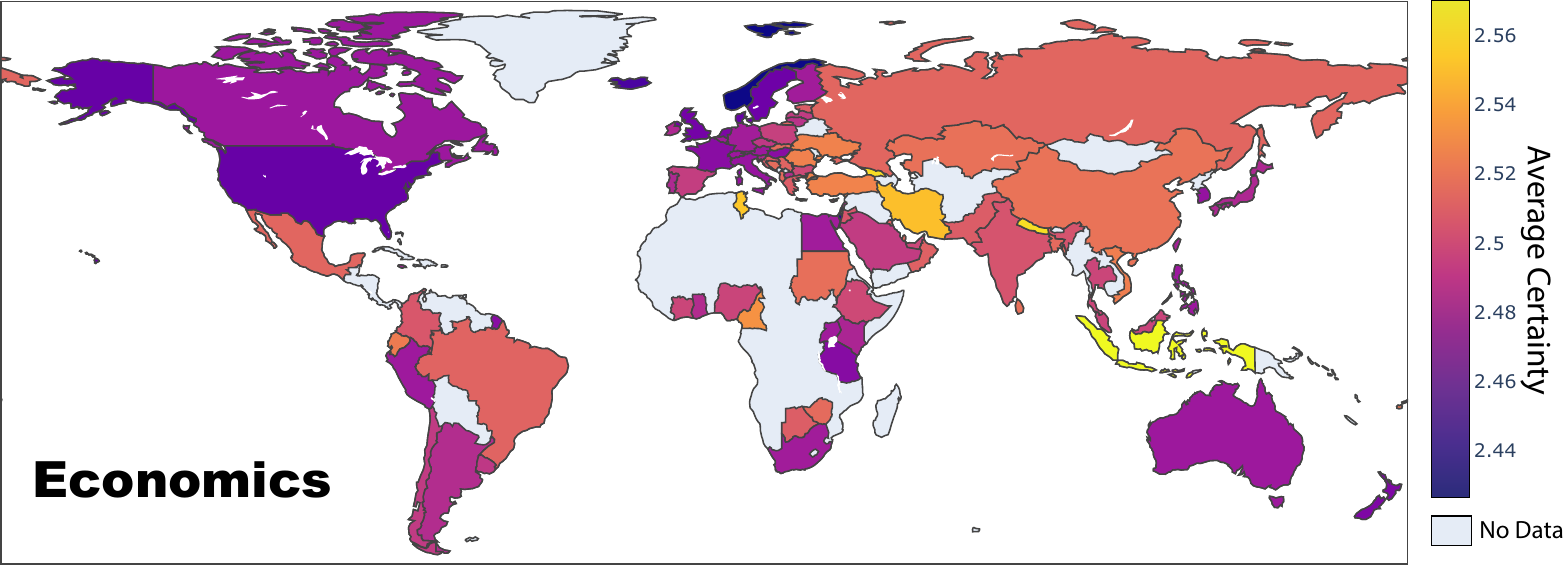}
\end{adjustwidth}
\end{figure}

\end{adjustwidth}

\end{document}